\documentclass[aps,prl,twocolumn,groupedaddress]{revtex4} 

\newcounter{col}
\usepackage[pdftex]{graphicx}
\usepackage{subfigure}
\usepackage{verbatim}
\usepackage{amsmath}
\usepackage{amssymb}
\usepackage{ifthen}

\newcommand{\beq}{\begin{equation}}
\newcommand{\eeq}{\end{equation}}
\newcommand{\beqn}{\begin{eqnarray}}
\newcommand{\eeqn}{\end{eqnarray}}
\newcommand{\avg}[1]{\langle{#1}\rangle}

\begin{document}

\title{Quantifying evolvability in small biological networks}

\author{Andrew Mugler} \email{ajm2121@columbia.edu}
\affiliation{Department of Physics, Columbia University, New York, NY
  10027}

\author{Etay Ziv}\email{ez87@columbia.edu} \affiliation{College of
  Physicians and Surgeons, Columbia University, New York, NY 10027}

\author{Ilya Nemenman}\email{nemenman@lanl.gov} \affiliation{Computer,
  Computational and Statistical Sciences Division, and Center for
  Nonlinear Studies, Los Alamos National Laboratory, Los Alamos, NM
  87545}

\author{Chris H. Wiggins}\email{chris.wiggins@columbia.edu}
\affiliation{Department of Applied Physics and Applied Mathematics,
  Center for Computational Biology and Bioinformatics, Columbia
  University, New York, NY 10027}

\date{\today}

\begin{abstract}
  We introduce a quantitative measure of the capacity of a small
  biological network to evolve. We apply our measure to a stochastic
  description of the experimental setup of Guet et al.\ ({\em Science}
  {\bf 296}:1466, 2002), treating chemical inducers as functional
  inputs to biochemical networks and the expression of a reporter gene
  as the functional output. We take an information-theoretic approach,
  allowing the system to set parameters that optimize signal
  processing ability, thus enumerating each network's highest-fidelity
  functions. We find that all networks studied are highly evolvable by
  our measure, meaning that change in function has little dependence
  on change in parameters.  Moreover, we find that each network's
  functions are connected by paths in the parameter space along which
  information is not significantly lowered, meaning a network may
  continuously change its functionality without losing it along the way.
  This property further underscores the evolvability of the networks.
\end{abstract}

\maketitle

Many signals in cells are processed using a network of interacting
genes: exogenous signals affect expression of genes coding for
transcription factor proteins, which in turn regulate the expression of
other genes.  Although early works have suggested that the
connectivity of such regulatory networks dictates their function
\cite{ShenOrr,Mangan,Kollmann}, recent studies offer evidence that a
network with fixed connectivity can change its function simply by
varying its biochemical parameters \cite{Wall,Ziv,Voigt,Guet}.  The
diversity of a network's achievable functions and the ease with which
it can realize them are central to its capacity to evolve
epigenetically, without slow and costly modifications to the genetic
code, and thus central to the evolutionary capacity of the organism as a
whole.

The evolvability of a regulatory network has been a topic of much
discussion in recent literature
\cite{Voigt,Ptashne,Kitano,Buchler,Braunewell,Kashtan}, but little has
been done to quantify the concept in a principled way.  Here we
propose a quantitative measure of network evolvability, and we apply
it to a set of small regulatory networks, such that a principled
comparison can be made across networks.  Networks are taken from the
experimental setup of Guet et al.\ \cite{Guet} and modeled
stochastically.  We find biochemical parameters that optimize the
information flow between a chemical ``input'' signal and a particular
``output'' gene, and we indeed find that a single network performs
different functions at different sets of optimal parameters.  We argue
that a more evolvable network will be able to access a richer
diversity of its functions with smaller changes in its parameters, and
as such we quantify evolvability using a measure of anti-correlation
between parametric and functional change.

We find that while there are small differences among networks'
evolvability scores, all are highly evolvable, meaning that the
magnitude of a functional change has little dependence on the
parametric change required to produce it.  Moreover, we find that
transitions among a network's optimally informative functions can be
made without significant loss of the input-output information along
the way.  By proposing and demonstrating a principled evolvability
measure, we reveal these features quantitatively; both features
suggest a high capacity of the studied regulatory networks to evolve.

\section{Methods}
First we briefly outline the methods used to develop a quantitative
measure of evolvability; each step is discussed in more detail in the
sections that follow.  The system of interest is a small (4-gene)
transcriptional regulatory network.  As in Guet et al.\ \cite{Guet},
we treat the presence or absence of chemical inducers (small molecules
that affect the efficacy of the transcription factors) as the inputs
to the network, and the expression of a particular gene as the output.
We use a stochastic model to find expression level distributions at a
steady-state, and we search for the model parameters which maximize
the mutual information between input and output.  We characterize the
function that the network performs by the order in which the output
distributions corresponding to each input state are encoded
\cite{Mugler}; a single network can perform different functions at
different parameter settings.  We define the evolvability of the
network as the ability to perform a diverse set of functions with only
small changes in parameters, and we quantify evolvability accordingly
using a measure of anti-correlation between pair-wise function distance
and parameter distance.

\subsection{Model}
Following the experimental setup of Guet et al.\ \cite{Guet}, we study
all networks that can be built out of three genes $A$, $B$, and $C$,
in which each gene is regulated by one other gene, and regulation
edges can be up-regulating or down-regulating.  Additionally, as in
the experiment, gene $C$ down-regulates a ``reporter'' gene $G$ (e.g.,
GFP), whose expression we treat as the functional output of the
network.  This yields a total of 24 networks, as shown on the
horizontal axis of Figure \ref{Erank}.  Also in analogy to the
experiment, the efficacy of each transcription factor can be inhibited
by the presence of a chemical inducer $s$, a small molecule that binds
to the transcription factor and lowers its affinity for its binding
site.  The presence or absence of the chemical inducers $s_A$, $s_B$,
and $s_C$ corresponding to each transcription factor $A$, $B$, and $C$
define the functional input state of the network.  The inhibitory
effect of each inducer is illustrated for an example network in the
top panel of Figure \ref{evo}A, and the eight possible input states
$i$, determined by the presence or absence of the inducers, are listed
in the bottom panel.

For a typical regulatory network inside a cell, intrinsic noise
arising from fluctuations in the small numbers of species
\cite{Elowitz2,Thattai2} is the primary factor
limiting transmission of information from a chemical input to a
genetic output \cite{Ziv,Acar,Pedraza,Mugler}. This
observation has two important consequences for modeling: (a) a
realistic model should capture not just mean protein concentrations,
but probability distributions over numbers of proteins
\cite{Shahrezaei,Hornos}, and (b) the most biologically relevant model
parameters will retain an optimal flow of information in the presence
of this noise \cite{Ziv,Tkacik,Doan}.

The first consequence is most fully addressed by solving the chemical
master equation \cite{vanKampen}, which describes the time-evolution
of the joint probability distribution for the numbers of all molecules
in the system given the elementary reactions (which are ultimately a
function of the network topology). For our systems, the master
equation is not analytically solvable. Progress can be made either by
Monte-Carlo simulation of the master equation \cite{Gillespie}, or by
approximating the master equation, e.g.\ with the linear-noise
approximation (LNA) \cite{Paulsson,Elf2,vanKampen,Ziv,Mugler}.
Since it does not rely on sampling, the LNA is much more
computationally efficient (and thus more amenable to a search for high-fidelity model parameters), and in previous work \cite{Ziv} we found
the distributions obtained via the LNA were practically
indistinguishable from those obtained via stochastic simulation for
copy numbers above $10-20$.

In the LNA, the reaction rates in the master
equation are linearized, and the steady-state solution is a
multidimensional Gaussian distribution \cite{vanKampen,Ziv,Mugler}.
The individual species' marginal distributions are thus described at the level of Gaussian fluctuations, with means given
by the steady-state solution to the deterministic dynamical system
describing mean protein numbers. Mean expression has been modeled with
remarkable success by combining transcription and translation into one
step \cite{Elowitz,Gardner,Hasty}, and accordingly, for each of our
networks, we use the following dynamical system in which species are
directly coupled to one another,
\beq
\label{dyn}
\frac{dX_j}{dt} = \alpha_j(X_{\pi_j}/s_{\pi_j})-R_jX_j,
\eeq
where the $X_j\in\{A,B,C,G\}$ are the expression levels (in units of
proteins per cell) of the four genes, the $R_j$ are degradation rates,
and the $\alpha_j$ are production rates which depend on the expression
level $X_{\pi_j}$ and a scale factor $s_{\pi_j}$ of the parent $\pi_j$
of gene $j$ (where the parent-children connectivity is determined by
the network topology). The scale factors,
$s_{\pi_j}\in\{s_A,s_B,s_C\}\ge1$, incorporate the inhibitory effect
of each chemical inducer (when present) by reducing the effective
transcription factor concentration; when an inducer is absent, its
scale factor is set to 1. Regulation is modeled using Hill functions,
\beq
\label{hill}
\alpha_j(x) = 
\begin{cases}
	a_0 + a_j\frac{x^n}{x^n+(K_j)^n} & {\rm up-regulating}, \\
	a_0 + a_j\frac{(K_j)^n}{x^n+(K_j)^n} & {\rm down-regulating},
\end{cases}
\eeq where $a_0$ (kept the same for all genes) is the basal production
rate, $a_0+a_j$ is the maximal production rate, $K_j$ is the binding
constant or the protein number at which production is half-maximal,
and $n$ is the cooperativity, which we set to 2. Note from Eqns.\
(\ref{dyn}-\ref{hill}) that increasing the scale factor can be
equivalently interpreted as increasing the effective binding constant
or lowering the binding affinity. Steady states of Eqn.\ (\ref{dyn})
are found by solving (using MATLAB's {\tt roots}) the polynomial
equations that result from setting the left-hand side to zero, and
keeping only those solutions for which the Jacobian matrix $J$ of
Eqn.\ (\ref{dyn}) has eigenvalues whose real parts are all negative.

The variances of the marginal distributions are the diagonal entries in the covariance matrix $\Xi$, which under the LNA satisfies a Lyapunov equation,
\beq
\label{lyap}
J\Xi+\Xi J^T+D=0, \eeq where $D$ is a diagonal matrix with, for the
system in Eqn.\ (\ref{dyn}), the $j$th entry equal to $\alpha_j(X_{\pi_j}/s_{\pi_j})+R_jX_j$.  Eqn.\ (\ref{lyap}) is solved using a standard Lyapunov solver
(MATLAB's {\tt lyap}). Since the steady-state distributions are
Gaussian under the LNA, the solution is fully specified by the means
and the variances. The distributions of particular functional
importance are $P(G|i)$, the probability of expressing $G$ reporter
proteins per cell given that the chemical inducers are in state $i$.

To address the second consequence, that biologically relevant
solutions often optimize information flow in the presence of intrinsic
noise \cite{Tkacik,Doan}, as in previous work \cite{Ziv} we allow
the system to set parameters that maximize the mutual information $I$
\cite{Shannon} between input state $i$ and output expression $G$,
where
\beqn
I &=& \sum_i \int dG \, P(i,G) \log_2 \frac{P(i,G)}{P(i)P(G)}\\
\label{I}
&=& \frac{1}{|i|}\sum_{i=1}^{|i|} \int dG \, p(G|i) \log_2
\frac{|i|P(G|i)}{\sum_{i'=1}^{|i|}P(G|i')}. \eeqn Here $I$ is measured
in bits, and the second step uses $P(i,G)=P(G|i)P(i)$,
$P(G)=\sum_{i'}P(i',G)$, and an assertion that each input state occurs
with equal likelihood (i.e., $P(i)=1/|i|$, where $|i|=8$ is the number
of input states) to write $I$ entirely in terms of the model solutions
$P(G|i)$.

Two computationally trivial ways for the system to maximize $I$ are
(a) to use an unbounded number of reporter proteins $G$ to encode the
signal, and (b) to set degradation rates such that $G$ responds on a
timescale much longer than that of the upstream genes (called a
``stiff'' system), which has the effect of averaging out the upstream
noise. In contrast, in cells, protein production requires energy, which sets a limit on the number of proteins that a cell can produce, and most protein
degradation rates are comparable. Therefore we seek model parameters
$\vec{\theta}^*$ that optimize a constrained objective function, \beq
\vec{\theta}^* = {\rm
  argmax}_{\vec{\theta}}\left[I-\lambda\avg{X_j}-\gamma\avg{R_{\pi_j}}/R_G\right],
\eeq where the constants $\lambda$ and $\gamma$ are a metabolic cost
and a constraint against stiffness, respectively, the average
$\avg{X_j}$ is taken over all genes, and the average $\avg{R_{\pi_j}}$
is taken over upstream genes $A$, $B$, and $C$. Optimization is
performed using a simplex algorithm (MATLAB's {\tt fminsearch}) in a
$15$-dimensional parameter space, as 
\ifthenelse{\value{col} = 1}{
\beq 
\vec{\theta} =
\{a_0,a_A,a_B,a_C,a_G,K_A,K_B,K_C,K_G,R_A,R_B,R_C,s_A,s_B,s_C\}
\eeq
}{
\beqn
\vec{\theta} &=&
\{a_0,a_A,a_B,a_C,a_G,K_A,K_B,K_C,K_G,\nonumber\\
&&R_A,R_B,R_C,s_A,s_B,s_C\}
\eeqn
}
($R_G$ was fixed at $4\times10^{-4}s^{-1}$ to set a biologically
realistic degradation rate scale).

Varying initial parameters yields many local optima $\vec{\theta}^*$
at which the input signal may be encoded differently in the output
distributions $P(G|i)$. For example, two optimally informative
solutions are shown in Figure \ref{evo}B for the network in Figure
\ref{evo}A. Intuitively, maximizing mutual information has resulted in
sets of distributions that are well separated, such that knowledge of
the output $G$ would leave little ambiguity about the original input
state $i$. We point out, however, that the ordering of the output
distributions is different between the two solutions, meaning that the
network is performing two different functions at two different points
in parameter space. The relationship between diversity of such
functions and exploration of parameters is crucial to the discussion
of evolvability; in the next section we develop a quantitative measure
of evolvability in the context of this system.

\subsection{Quantifying evolvability}
As seen in several experimental and numerical studies
\cite{Wall,Ziv,Voigt,Guet}, and in data from the model described
above, a single regulatory network can perform different functions
simply by varying its biochemical parameters. Intuitively, a network
should be deemed more evolvable if it is able to access a richer
diversity of its functions with smaller changes in its parameters.
Quantification of this concept requires definitions of both parametric
and functional change.

As in Barkai et al.\ \cite{Barkai}, we characterize the magnitude of
the parametric change in going from one model solution to another by
calculating fold-changes in the model parameters. Specifically, we
define a parameter distance $\Delta\theta$ between two solutions as
the Euclidean distance in the logs of the parameters,
\beq
\label{dq}
\Delta\theta = |\log_2\vec{\theta}^*_1-\log_2\vec{\theta}^*_2| =
\sqrt{\sum_{k=1}^{|k|}\left[\log_2(\theta^*_{1,k}/\theta^*_{2,k})\right]^2},
\eeq where $|k|=15$ is the number of parameters. Under this
definition, equal fold-changes in each parameter constitute equal
contributions to $\Delta\theta$ (for scale, the doubling of one
parameter corresponds to $\Delta\theta=1$).

\begin{figure*} \centering
  \begin{minipage}[c]{0.15\textwidth}
    \includegraphics[width=\textwidth]{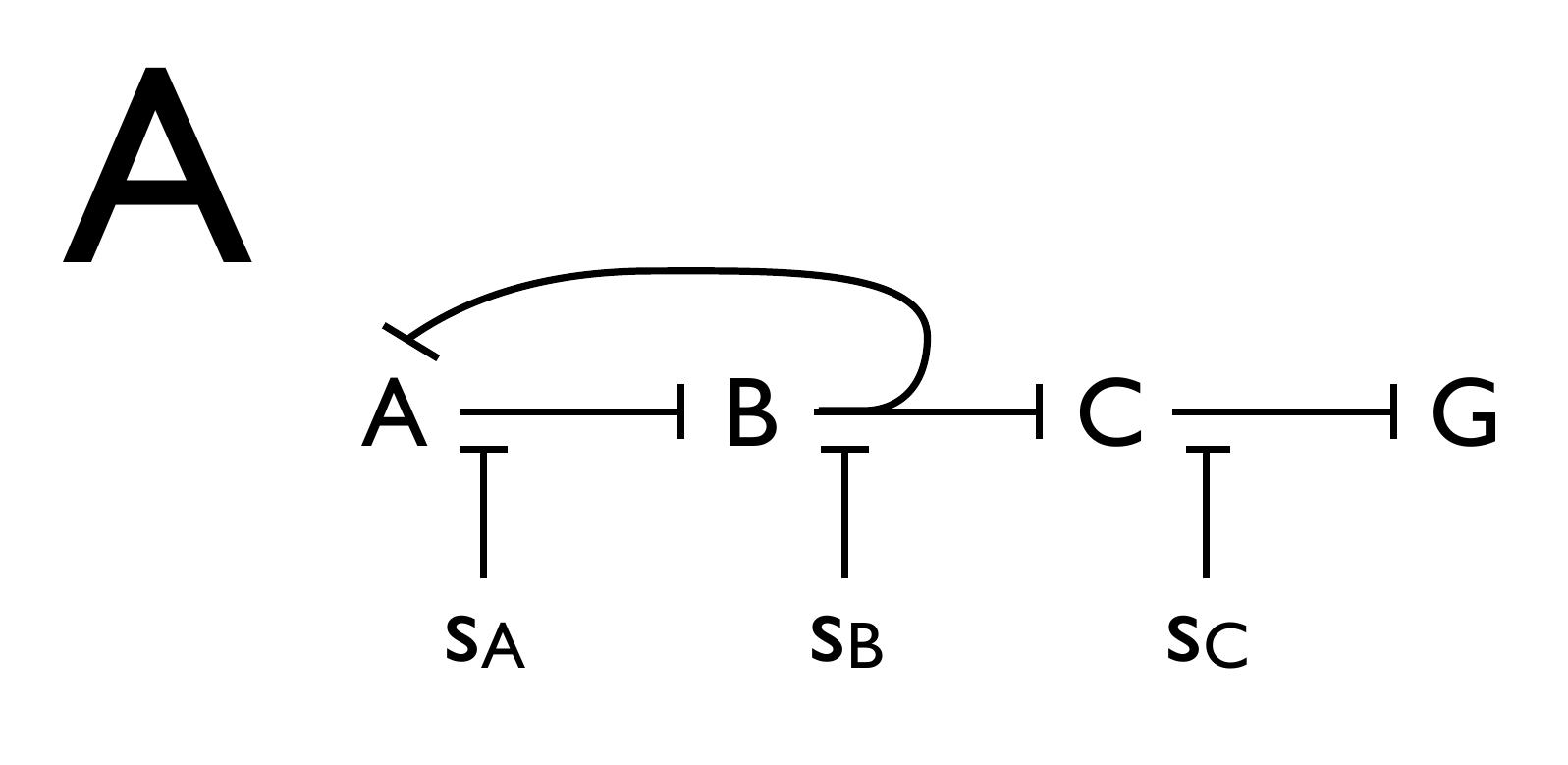}\\
    \includegraphics[width=\textwidth]{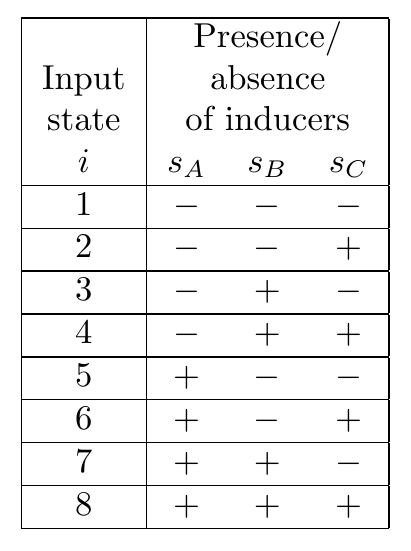} \end{minipage}
  $\qquad$ \begin{minipage}[c]{0.35\textwidth}
    \includegraphics[width=\textwidth]{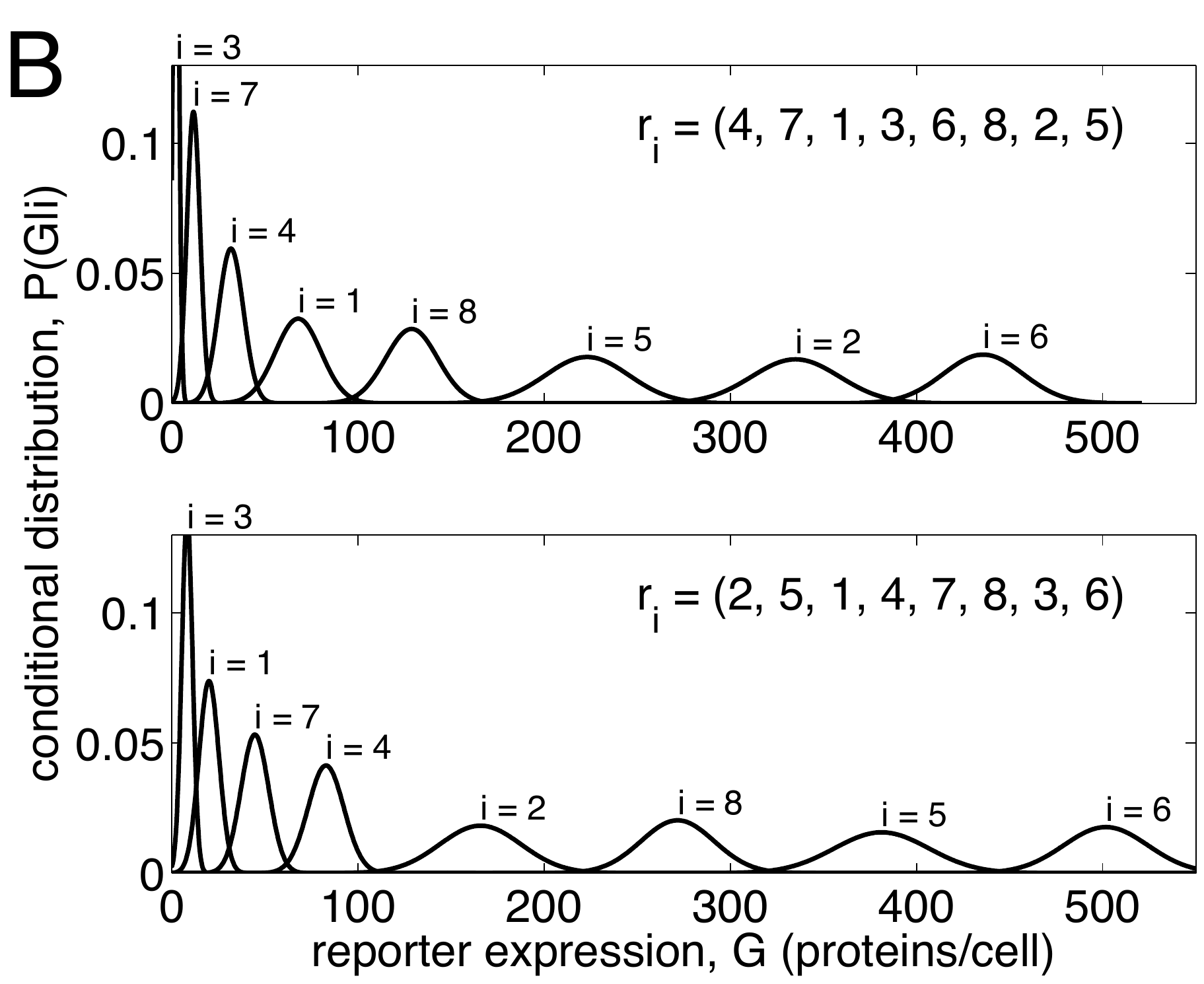} \end{minipage} $\qquad$
  \begin{minipage}[c]{0.35\textwidth}
    \includegraphics[width=\textwidth]{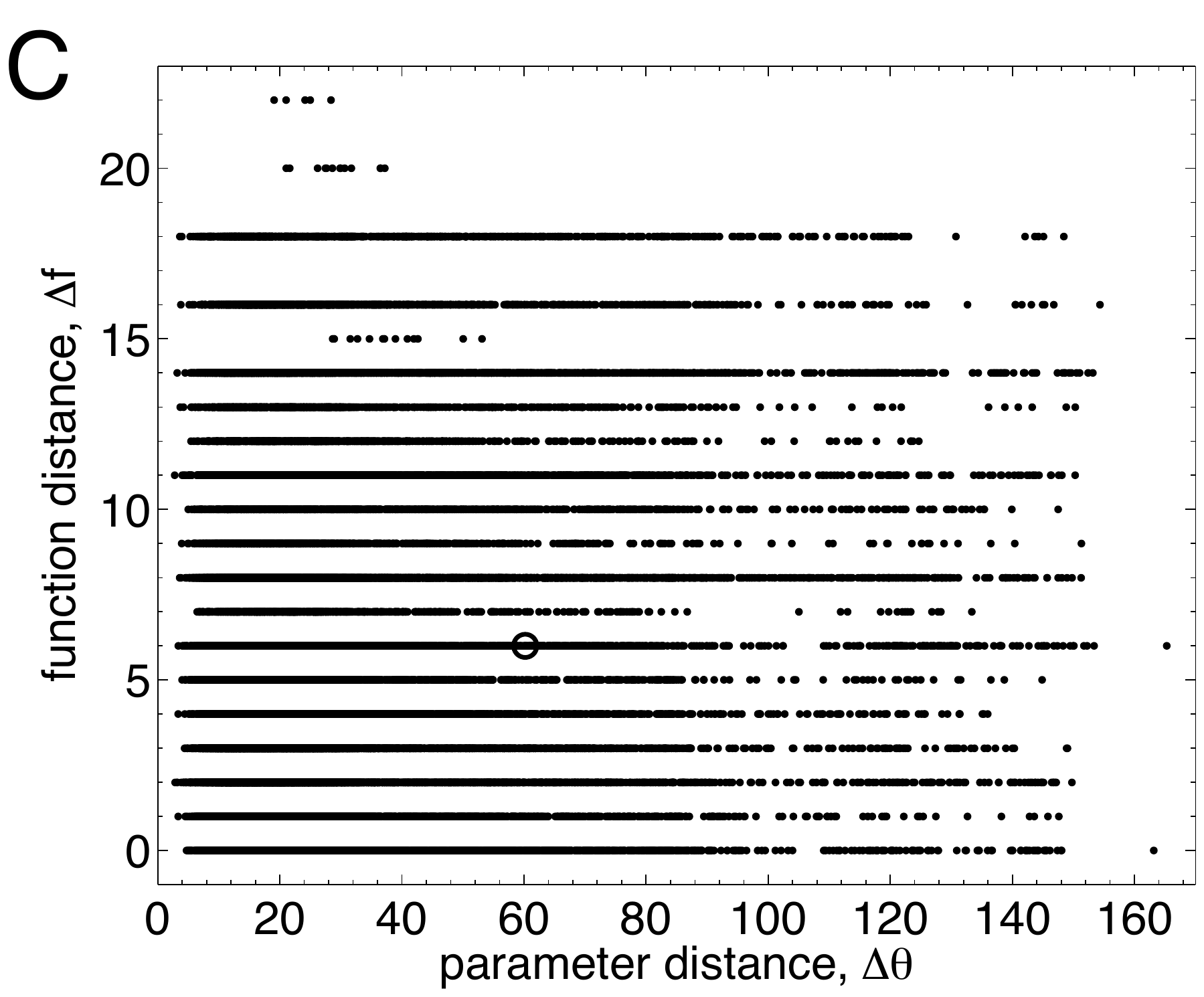}
  \end{minipage}
  \linespread{1} \caption{Defining evolvability. {\bf A
      Top:} a sample regulatory network (see Figure \ref{Erank} for
    diagrams of all 24 networks studied). $A$, $B$, and $C$ are genes
    whose transcription factors regulate each other's expression
    according to the given network topology, and $G$ is a ``reporter''
    gene, such as GFP. Sharp arrows indicate up-regulation, while
    blunt arrows indicate down-regulation (all arrows are blunt in
    this network). $s_A$, $s_B$, and $s_C$ are chemical inducers that
    reduce the efficacy of the corresponding transcription factors.
    {\bf Bottom:} Table showing the 8 input states $i$ that are
    defined by the presence or absence of each chemical inducer in the
    cell ($+$ indicates presence, $-$ indicates absence). In the model,
    the $s_A$, $s_B$, and $s_C$ are scale factors that are free
    parameters (greater than 1, to effect an interference with
    transcriptional regulation) if the inducer is present, and are set
    to 1 if the inducer is absent. {\bf B:} Two maximally informative
    functions performed by the sample network at two different
    parameter settings. Function is characterized by the order of the
    output distributions $P(G|i)$, the probability of expressing $G$
    proteins per cell given that the system is in input state $i$.
    Specifically, the function is quantified by the vector $\vec{r}$
    of ranks of the $P(G|i)$, as shown for each function in the upper
    right corner. For example, in the top function, the first output
    distribution ($i=1$) is ranked $4$th, the second ($i=2$) is ranked
    $7$th, the third ($i=3$) is ranked $1$st, and so on, so the rank
    vector is $\vec{r} = (4,7,1,\dots)$. {\bf C:} Plot of function
    distance $\Delta f$ [Eqn.\ (\ref{df})] vs.\ parameter distance
    $\Delta \theta$ [Eqn.\ (\ref{dq})] for all pairs of maximally
    informative model solutions ($340$ solutions were used for this
    network). Function distance is scaled such that the swapping of
    two adjacent output states from solution one to solution two gives
    $\Delta f=1$, and parameter distance is scaled such that the
    doubling of one parameter from solution one to solution two gives
    $\Delta\theta=1$. The point corresponding to the pair of functions
    in B is circled. The evolvability score for this network,
    calculated from these data via Eqn.\ (\ref{E}), is
    $E=0.482\pm0.001$.}
\label{evo}
\end{figure*}

As in previous work \cite{Mugler} and in the original experiment of
Guet et al.\ \cite{Guet}, we define the function of a network
analogously to logic in electrical circuits (AND, OR, XOR, etc.), in
which the function is determined by the magnitude of the output's
response to each input state (for example, with two inputs, AND would
be defined by a ``high'' output in response to the $[++]$ state, and a
``low'' output in response to the $[--]$, $[-+]$, and $[+-]$ states).
Since, in our setup, optimizing information produces well-separated
output distributions $P(G|i)$ (see Figure \ref{evo}B), we extend this
idea beyond a simple ``high'' or ``low'' output classification, and
characterize function by the order of the $P(G|i)$. Specifically, we
record a vector $\vec{r}$ of ranks of the $P(G|i)$; for example, in
the top panel of Figure \ref{evo}B, the first output distribution
($i=1$) is ranked $4$th, the second ($i=2$) is ranked $7$th, the third
($i=3$) is ranked $1$st, and so on, so the rank vector is $\vec{r} =
(4,7,1,\dots)$. We then define the function distance $\Delta f$
between two solutions in terms of the vector distance between their
rank vectors,
\beq
\label{df}
\Delta f = \frac{1}{2}|\vec{r}_1-\vec{r}_2|^2 =
\frac{1}{2}\sum_{i=1}^{|i|}(r_{1,i}-r_{2,i})^2 \eeq (for scale, the
swapping of two adjacent output distributions corresponds to $\Delta
f=1$) \footnote{In networks in which the overall sign of the feedback
  cycle is negative, there can exist parameter values that support
  multiple stable fixed points. This would correspond to one or more
  of the output distributions being multimodal. Since we effectively
  minimize overlap of output states by optimizing information
  transmission, such solutions are rare (13\% occurrence in all
  negative-feedback networks). When they do occur, we equally weight
  each fixed point in constructing the multimodal Gaussian output, and
  continue to define $\vec{r}$ by the ranks of the means of the output
  distributions.}. Other function distances, including other
permutation distances between the rank vectors, and a continuous
distance measure defined by averaging the Jensen-Shannon divergence
\cite{Lin} between corresponding output distributions in the solution
pair, produced similar results, as discussed in the Results section.

It is now clear that, if a network is better able to explore its
function set with smaller changes in its parameters (i.e., is more
evolvable by our definition), then it will exhibit less correlation
between $\Delta f$ and $\Delta\theta$ than other networks. Therefore
we define an evolvability score $E$ for a given network as a measure
of anti-correlation between $\Delta f$ and $\Delta\theta$, calculated
for every pair of its optimal model solutions \footnote{If two
  solutions from the same local information maximum are treated as
  distinct, they will have the same function but (slightly) different
  parameters; this will artificially lower $E$. To correct for this
  effect, we merge (at their mean parameter location) nearest
  neighbors whose functions are the same until all nearest neighbors
  have different functions. This procedure reduced networks' solution
  sets by at most $\sim$$10 \%$.}. Specifically,
\beq
\label{E}
E = 1-(\tau+1)/2, \eeq where $\tau$ is Kendall's tau \cite{Kendall}, a
nonparametric measure of correlation between all pairwise $\Delta f$
and $\Delta\theta$; we rescale $\tau$ such that $0<E<1$ and take its
complement to obtain an anti-correlation. Using a nonparametric
correlation statistic has the advantage that our evolvability measure
remains invariant upon any monotonic rescaling in the definitions of
either $\Delta\theta$ or $\Delta f$. Additionally, we note that $E$
can be thought of as the probability that a pair of solutions drawn at
random have a larger $\Delta f$ than another pair given that the first
pair had a smaller $\Delta\theta$, or as the fraction of discordant
pairs of $(\Delta\theta,\Delta f)$ data points \footnote{Many sources
  (including MATLAB's built-in {\tt corr}) use an adjustment to the
  calculation of $\tau$ in the case of tied data (see e.g.\
  \cite{Kendall2}). In keeping with the interpretation of our
  statistic as a probability, we do not introduce an adjustment; we
  simply count each tied pair as neither concordant nor discordant
  (i.e.\ if, for example, in computing the fraction of concordant
  pairs, we assigned each concordant pair a $1$ and each discordant
  pair a $0$, a tied pair would count as $0.5$).}.

Function distance $\Delta f$ vs.\ parameter distance $\Delta\theta$
for all pairs of model solutions is plotted in Figure \ref{evo}C for
the example network in Figure \ref{evo}A. The evolvability score
calculated from these data is $E=0.482$ which, since there is little
correlation (or anti-correlation) between $\Delta f$ and $\Delta\theta$
in this case, is near the middle value $E=0.5$.

We obtain a fairer estimate of $E$ and an estimate of its error by subsampling.  Specifically, in the spirit of Strong et al.\ \cite{Strong}, we compute the mean $\bar{E}$ and standard error $\delta E$ in $E$ values calculated on randomly drawn subsets of a given size $n$ (from the full data set of size $N$).  We then repeat for various $n$, plot $\bar{E}\pm\delta E$ vs.\ $N/n$, and fit with a line (all plots generated were roughly linear).  The value and uncertainty of the $N/n=0$ intercept give an estimate of $E$, extrapolated to infinite data, and a measure of sampling error, respectively.  The sampling error estimated in this way for the data in Figure \ref{evo}C is $0.001$.

\section{Results}
\subsection{All networks studied are evolvable}
\begin{figure} \centering
\ifthenelse{\value{col} = 1}{
  \includegraphics[width=.8\textwidth]{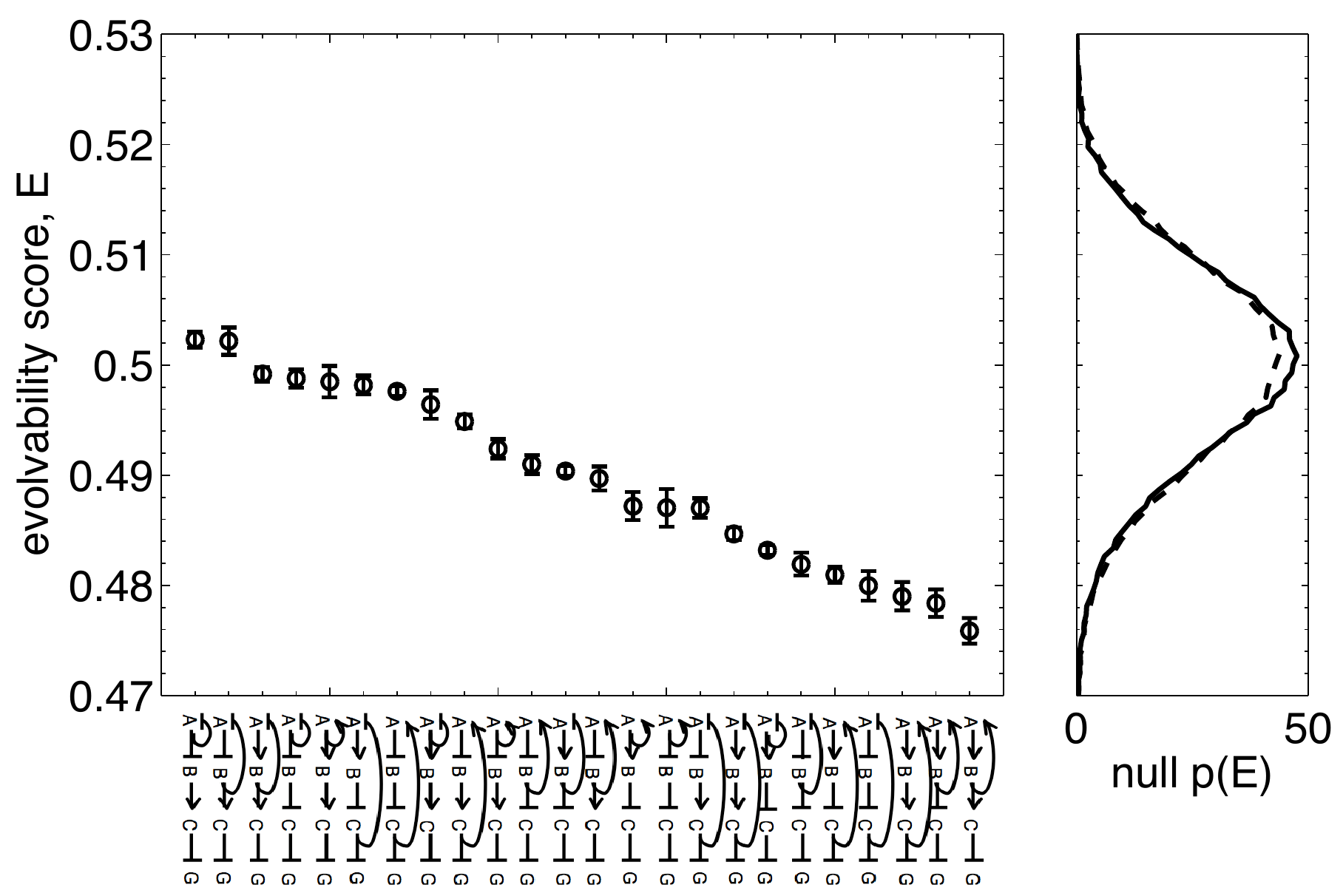}
}{
  \includegraphics[width=.48\textwidth]{Epvals.pdf}
}
  \linespread{1}
  \caption{{\bf Left:} Evolvability scores $E$ for all 24 regulatory
    networks studied. Networks are shown along the horizontal axis,
    ranked by $E$ (sharp arrows denote up-regulation, and blunt arrows
    denote down-regulation). $E$ values are calculated via Eqn.\
    (\ref{E}), with error bars showing the sampling error, calculated as
    described in the text. {\bf Right:} Two null distributions
    generated according to the null hypothesis that the function distance
    is independent of the parameter distance. The solid line is the
    distribution of $E$ scores calculated from solution sets in which
    the locations in parameter space were held fixed, and the function
    assignments were randomly permuted. The dotted line is the
    distribution of $E$ scores calculated from solution sets in which
    the locations in parameter space were held fixed, and the function
    assignments were drawn randomly from the set of possible functions
    for the given network. Both distributions are averages over the
    individual distributions for each network, as there was no
    correlation between the means or variances of the individual
    distributions and the networks' $E$ scores.}
\label{Erank}
\end{figure}

Using the methods described above, between $200$ and $500$ optimally
informative model solutions were obtained, and an evolvability score
$E$ was calculated for each of the 24 networks shown on the horizontal
axis of Fig.~\ref{Erank}. The constraints were set to $\lambda=0.01$
or $0.005$, for an average protein count of $\sim$$100-200$, and
$\gamma=0.001$, allowing a maximum of about 3 orders of magnitude
between upstream and reporter degradation rates. Solutions with mutual
information values below $I=2$ bits were discarded as not transmitting
high enough information (for scale, a solution with perfectly
overlapping output distributions would have $I=0$ bits, and a solution
with 8 perfectly non-overlapping output states would have $I=3$ bits).

Networks' evolvability scores are shown in Fig.~\ref{Erank}. All 24
networks had $E$ values within $5\%$ of 0.5 (recall that $E$ is
bounded by $0\le E \le 1$), which means that, in all cases, there is
little correlation between change in function and change in
parameters, suggesting that all networks studied are evolvable. Using
other function distances, including other permutation distances
between the rank vectors, and a continuous distance measure defined by
averaging the Jensen-Shannon divergence \cite{Lin} between
corresponding output distributions in the solution pair, produced
similar results: $E$ scores were very near $0.5$, indicating little
correlation between functional and parametric distances.

The claim that function has little dependence on parameters can be
tested more rigorously by comparison with a null hypothesis. The null
hypothesis that function is independent of parameters was implemented
in two ways. First, given each network's solution set, locations of
solutions in parameter space were kept the same, but the functions
associated with each solution were randomly permuted. Second,
locations of solutions in parameter space were again kept the same,
but functions were drawn randomly from the set of possible functions
for each network \footnote{Not all $8!$ rankings of the output
  distributions are allowed functions for a given network. As shown in
  previous work \cite{Mugler}, the topology of the network constrains
  the set of possible steady-state functions. Specifically, since each
  gene is regulated by one other gene, allowed functions are
  ``direct'' functions: those in which the output distribution
  responds to a change in inducer concentration according to the
  direct path from inducer to reporter (i.e., ignoring feedback
  pathways). For example, for the network in Fig.~\ref{evo}A, in
  going from state $[---]$ ($i$=1) to $[-+-]$ ($i$=3), $s_B$
  increases; the direct path from $s_B$ to $G$ consists of a
  repression--repression--repression chain, which is net repressive,
  so the output distribution must decrease (as it does in both panels
  of Fig.~\ref{evo}B). With $3$ inducers, there are $48$ direct
  functions for each network; this is the set from which functions are
  randomly drawn in the second implementation of the null
  hypothesis.}. In each case, the function reassignment was performed
many times, and the $E$ value was computed each time to produce a
distribution of null $E$ scores. There was no correlation between the
means or variances of the networks' null distributions and their
actual $E$ scores, so the individual null distributions were averaged
across networks. Averaged null distributions from each of the two
implementations are qualitatively similar, and both are shown in
Figure \ref{Erank}. All networks' actual $E$ values lie well within
both null distributions (the smallest $p$-value is $0.023$, and, with 24 networks, we expect at least one
to attain a $p$-value lower than $1/24=0.04$ simply by chance). This means that none of the networks' solution sets
significantly differ from a set in which the function performed is
independent of the setting of the parameters.

Even though all $E$ values lie within the null distribution, only two lie above the null mean of $0.5$; the probability of this happening by chance is $2\times10^{-5}$.  For a network with $E$ much larger than $0.5$, the parameter and the functional
distances would be anti-correlated, and the network function would
evolve dramatically with very small parameter changes. Thus the vast
majority of the networks studied show a statistically significant, yet
unintuitively small, positive correlation among the functional and the
parametric distance.

Despite the fact that the $E$ values lie in a narrow range, sampling errors are small (see Fig.~\ref{Erank}), meaning
that the networks can be ranked with some confidence according to
their evolvability. We asked statistically whether this ranking was
correlated with any topological features of the network, including the
sign of the regulation of each gene, the length and net sign of the
feedback cycle, and the total number of activators and repressors in
the network, both in and out of the cycle. Correlation was tested for
features with categorical values using a Wilcoxon rank-sum test
\cite{Wilcoxon,Mann} (for two categories) or a Kruskal-Wallis $H$-test
\cite{Kruskal} (for more than two categories), and for features with
real values using Kendall's $\tau$ \cite{Kendall}. No topological
feature significantly correlated with $E$. The lowest $p$-value was
$0.04$, and, since many correlations were tested for at once, a
Bonferoni correction \cite{Salkind} showed that the likelihood of
obtaining a $p$-value this low simply by chance was $0.33$. Thus we
identified no topological aspect that significantly imparted higher or
lower evolvability to the networks.

\subsection{Changing functions without losing functionality}
As described in the previous section, we have found that the networks studied
organize their optimally informative solutions in parameter space in
such a way that change in function is largely independent of change in
parameters. We further demonstrate here that the networks can change
from one function to another in parameter space without significant
loss of the input-output information along the way. This further
underscores the evolvability of these networks, since it shows that
random steps in parameter space not only explore the full variety of a
network's functions, but do so without significant loss of fidelity.
In the context of electric logical circuits, such evolvability would
correspond to an ability to continuously change a logic gate from
performing one logical function to another while remaining a
functional gate in the interim.

\begin{figure}
  \includegraphics[width=.45\textwidth]{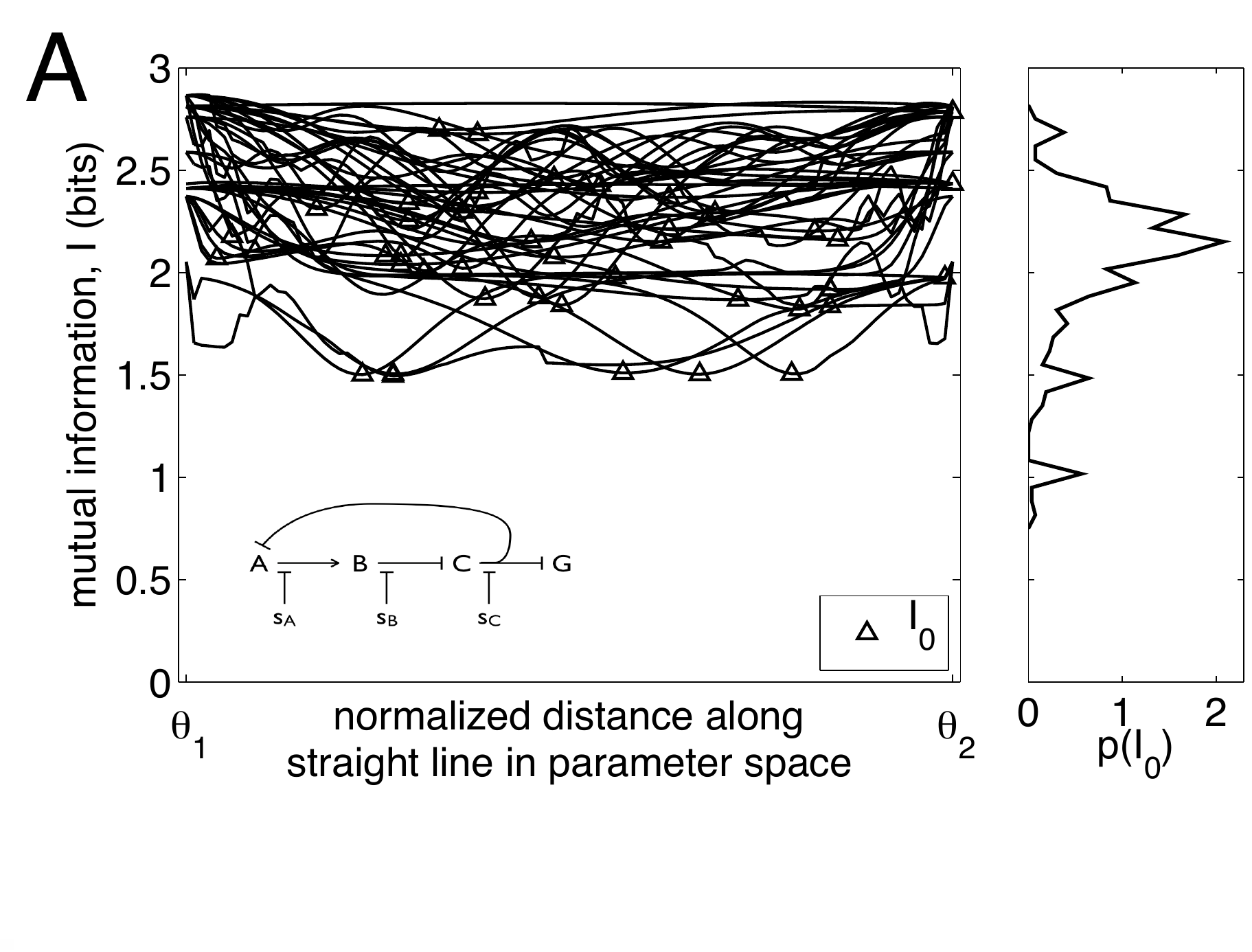}
  $\qquad$
  \includegraphics[width=.45\textwidth]{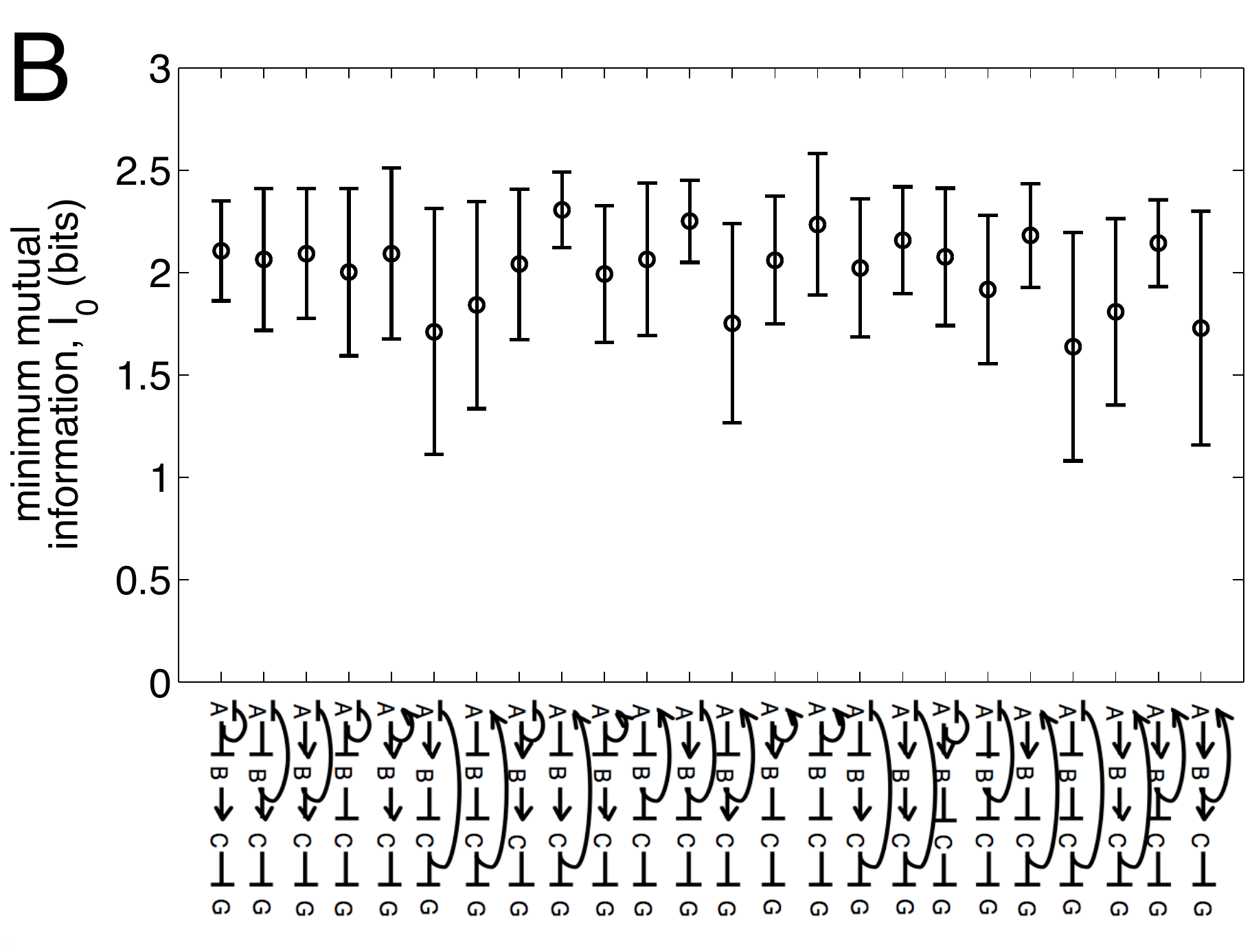}
  \linespread{1} \caption{Changing function without losing
    information. {\bf A Left:} Mutual information $I$ along
    straight-line paths in parameter space between pairs of $10$
    randomly chosen optimally informative model solutions for a
    particular network (inset). For each path, the starting and ending
    solution's locations in parameter space are denoted
    $\vec{\theta}_1$ and $\vec{\theta}_2$ respectively on the
    horizontal axis. The minimum mutual information $I_0$ along each
    path is marked with a triangle. A specific function is performed
    at each of the $10$ solutions (as characterized in Methods); $7$
    of the $10$ functions are unique. {\bf Right:} Distribution of
    $I_0$ values built from paths between $37$ randomly chosen
    solutions for the inset network, of which the $10$ solutions used
    for the left plot are a subset. {\bf B:} Means (circles) and
    standard deviations (error bars) of $I_0$ distributions like that
    in A (right), for all networks studied; 37 randomly chosen
    solutions were used to build each network's distribution. Networks
    are shown on the horizontal axis, in the same order as in Figure
    \ref{Erank}, i.e.\ ranked by evolvability score $E$.}
\label{paths}
\end{figure}

For each network, mutual information $I$ [Eqn.\ (\ref{I})] was
calculated along straight-line paths in parameter space between all
solutions pairs within a randomly chosen subset of its optimally
informative solutions. Examples of these paths are shown in Figure
\ref{paths}A, for $10$ solutions from the inset network. The solutions
at either end are local maxima in $I$, and the paths show the loss in
information capacity the network would suffer if it were to move from
one solution to the other along a straight line in parameter space.
Some information loss is unavoidable: changing function requires
reordering the output distributions (see Figure \ref{evo}B), which
means overlapping at least two of them at a time, and with $8$
distributions the shift of two distributions from fully separated to
fully overlapped incurs a minimum loss of at $0.25$ bits. Seven of the
$10$ functions corresponding to the $10$ solutions in Figure
\ref{paths}A are unique; at least $91\%$ of the plotted paths involve
a change in function.

Nonetheless, we find that the loss in information suffered in going
between optimal solutions is surprisingly minimal. The right panel of
Figure \ref{paths}A shows the distribution of minimal mutual
information values $I_0$ along the paths for the inset network, and
Figure \ref{paths}B shows the means and the standard deviations of
$I_0$ distributions for all networks. For only a few networks do a
significant portion of the paths drop below $1.5$ bits, and almost no
paths drop below $1$ bit. We note in passing that the networks in
Figure \ref{paths}B are shown as in Figure \ref{Erank}, i.e.\ ranked
by evolvability score $E$, and so Figure \ref{paths}B also
demonstrates that there is no significant correlation between $I_0$
and $E$.

We emphasize that Figure \ref{paths}B represents a lower bound on
minimum mutual information encountered in transitioning between
solutions. It is by no means necessary (and is most likely
biologically unrealistic) for a functional change to proceed via such
uniform changes in biochemical parameters. It is more likely that
there exist transition paths that are more optimal than the
straight-line paths, and that the most optimal $I_0$ distributions are
actually shifted higher in information than those generated here. Thus
it is quite nontrivial (and it is further testament to their
evolvability) that even along direct paths between optimal solutions
these networks in most cases do not drop below $1.5$ bits of
processing ability, considering that the solutions themselves operate
in the range of $\sim$$2-2.8$ bits. A network can be evolving and
functional at the same time.

\section{Discussion}
We have quantified the concept of evolvability in the context of
regulatory networks by introducing an interpretable measure, and by
probing the space of the networks' most informative functions. Our
measure is an anti-correlation between the amount of functional change
experienced by a network and the parametric change required to effect
it, such that more evolvable networks explore more diverse functions
with smaller variation in their biochemical parameters. We have fully
defined functional and parametric distances (as well as the
characterization of `function' itself) in the context of a stochastic
description of the experimental setup of Guet et al.\ \cite{Guet}, and
we have chosen a correlation measure that is invariant to monotonic
transformations in either definition.

We have found that all networks studied share the property that
functional change is largely independent of parametric change, meaning
that they are highly evolvable by our measure. This property holds for
several different definitions of function distance. This means that
high-information functions are not organized in parameter space in
such a way that similar functions are near each other; instead nearby
solutions are approximately as likely to be similar in function as
they are to be different in function.

Furthermore, we have found that all networks studied can transition
among their maximally informative functions without significant loss
of information in the process. Along straight-line paths in parameter
space between functions (with mutual information values in the range
$\sim$$2-2.8$ bits), mutual information remains above $\sim$$2$ bits
on average and very rarely drops below $1$ bit. Moreover, these values
represents a lower bound, since transition paths need not be straight.
This suggests that the networks can evolve without losing functionality
in the process, which resonates with the idea from evolutionary biology
that evolution happens not by crossing high fitness barriers (low-information solutions in our case), but by finding neutral paths \cite{Nimwegen}.

Ultimately we have uncovered two important properties of the
regulatory networks described by our model: (a) high-information
solutions do not cluster by function, and (b) transitions among
solutions are possible without significant loss of fidelity. Both of
these properties underscore the high evolvability of the networks
studied. It is possible that these properties are general
characteristics of a class of systems extending beyond small
transcriptional regulatory networks, particularly systems governed by
a large number of tunable parameters. However, we argue that these
properties are especially relevant here, as they are critical to a
quantitative description of the capacity of biological networks to
evolve.

\acknowledgements We are grateful to the organizers, participants, and
sponsors of The Second q-bio Conference in Santa Fe, New Mexico, where
a preliminary version of this work was presented. AM was supported by NSF Grant DGE-0742450.  IN was supported by
DOE under Contract No.\ DE-AC52-06NA25396 and by NSF Grant No.\
ECS-0425850.

\bibliographystyle{apsrev}
\bibliography{evo}

\end{document}